\documentclass[prl,showpacs,twocolumn,floatfix,superscriptaddress]{revtex4}
\usepackage{times,amsmath,amsfonts,amssymb,latexsym}
\usepackage{graphicx,epsf}
\newcommand{\be}{\begin{equation}}
\newcommand{\ee}{\end{equation}}
\newcommand{\ba}{\begin{eqnarray}}
\newcommand{\ea}{\end{eqnarray}}

\newcommand{\ignore}[1]{}
\def\CC{{\rm\kern.24em \vrule width.04em height1.46ex depth-.07ex
    \kern-.30em C}}
\def\P{{\rm I\kern-.25em P}}
\def\RR{{\rm
         \vrule width.04em height1.58ex depth-.0ex
         \kern-.04em R}}

\def\bbbc{{\mathchoice {\setbox0=\hbox{$\displaystyle\rm C$}\hbox{\hbox
to0pt{\kern0.4\wd0\vrule height0.9\ht0\hss}\box0}}
{\setbox0=\hbox{$\textstyle\rm C$}\hbox{\hbox
to0pt{\kern0.4\wd0\vrule height0.9\ht0\hss}\box0}}
{\setbox0=\hbox{$\scriptstyle\rm C$}\hbox{\hbox
to0pt{\kern0.4\wd0\vrule height0.9\ht0\hss}\box0}}
{\setbox0=\hbox{$\scriptscriptstyle\rm C$}\hbox{\hbox
to0pt{\kern0.4\wd0\vrule height0.9\ht0\hss}\box0}}}}
\def\bbbz{{\mathchoice {\hbox{$\sf\textstyle Z\kern-0.4em Z$}}
{\hbox{$\sf\textstyle Z\kern-0.4em Z$}}
{\hbox{$\sf\scriptstyle Z\kern-0.3em Z$}}
{\hbox{$\sf\scriptscriptstyle Z\kern-0.2em Z$}}}}

\begin{document}

\title{Lieb-Robinson bounds and the speed of light from topological order}
\author{Alioscia Hamma}
\affiliation{Perimeter Institute for Theoretical Physics\\
31 Caroline St. N, N2L 2Y5, Waterloo ON, Canada}
\affiliation{Massachusetts Institute of Technology, Research Laboratory of Electronics\\
77 Massachusetts Ave. Cambridge MA 02139}
\author{Fotini Markopoulou}
\affiliation{Perimeter Institute for Theoretical Physics\\
31 Caroline St. N, N2L 2Y5, Waterloo ON, Canada}
\affiliation{Massachusetts Institute of Technology, Research Laboratory of Electronics\\
77 Massachusetts Ave. Cambridge MA 02139}
\affiliation{Department of Physics, University of Waterloo \\ Waterloo, Ontario N2L 3G1, Canada}
\author{Isabeau Pr\'emont-Schwarz}
\affiliation{Perimeter Institute for Theoretical Physics\\
31 Caroline St. N, N2L 2Y5, Waterloo ON, Canada}
\affiliation{Massachusetts Institute of Technology, Research Laboratory of Electronics\\
77 Massachusetts Ave. Cambridge MA 02139}
\affiliation{Department of Physics, University of Waterloo \\ Waterloo, Ontario N2L 3G1, Canada}
\author{Simone Severini}
\affiliation{Institute for Quantum Computing and Department of Combinatorics \&
Optimization\\
University of Waterloo, 200 University Av. W, N2L 3G1, Waterloo ON, Canada}

\begin{abstract}
We apply the Lieb-Robinson bounds technique to find the maximum speed of
interaction in a spin model with topological order whose low-energy
effective theory describes light [see X.-G.~Wen, \prb {\bf 68}, 115413
(2003)]. The maximum speed of interactions is found in two dimensions is bounded from above less than $\sqrt{2} e$ times the speed of emerging light, giving a strong indication that light is indeed the maximum speed of interactions. This result does not rely on mean field theoretic
methods. In higher spatial dimensions, the Lieb-Robinson speed is
conjectured to increase linearly with the dimension itself. Implications for
the horizon problem in cosmology are discussed.
\end{abstract}

\pacs{11.15.-q, 71.10.-w, 05.50.+q}
\maketitle

\emph{Introduction.---} The \emph{principle of locality} is one of the most
fundamental ideas of modern physics. It states that every physical
system can be influenced only by those in its neighborhood. The concept of \emph{field} is the outcome of
taking this principle seriously: if object $A$ causes a change on
object $B$,  there must be changes involving the points in between. The
field is exactly what changes. In addition, if something is \textquotedblleft
happening\textquotedblright\ at all the intermediate points, then the
interaction between the objects must propagate with a finite speed.
Relativistic quantum mechanics is built by taking the locality
principle as a central feature. In
non-relativistic quantum mechanics the situation is more subtle:
signals can propagate at every speed and quantum correlations are non-local
in their nature.  One can, in fact, send information over any finite
distance in an arbitrary small time \cite{hastings}. However, the
amount of information that can be sent decreases exponentially with the
distance if the Hamiltonian of the system is the sum of local pieces.
Specifically there is an effective light cone resulting from a finite
maximum speed of the interactions in quantum systems. This is the essence of
the\emph{\ Lieb-Robinson bounds} \cite{lieb}. This notion have recently attracted
interest in the context of quantum information theory, condensed
matter physics, and the creation of topological order \cite{hastings, koma, eisert, sims}.

The concept of topological order is one of the most productive recent ideas
in condensed matter theory \cite{wentop}. It provides explanations for phases
of matter (for example, fractional quantum Hall liquids) that cannot be
described by the paradigm of local order parameters and symmetry breaking.
If local order parameters cannot describe such phenomena, then their order
could be of topological nature \cite{wentop}. Topological order gives rise
to a ground state degeneracy that depends on the topology of the system and
is robust against any local perturbations \cite{Wen:90}. Because of this
property, topologically ordered systems appear to be good candidates for
robust quantum memory and fault-tolerant quantum computation \cite{kitaev}.

Not only can topological order explain exotic phases of matter, but it
offers a whole new perspective to the problem of elementary particles. There
are particles that we regard as fundamental, like photons and fermions, and
other particles that can be interpreted as collective modes of a crystal.
For example, we can describe phonons in this way because of the symmetry of
the crystal. The understanding of the phases of matter provides an
explanation for the phonon and other gapless excitations. However,  one can also ask
whether  photons, electrons, or gravitons are  emergent phenomena too,
not elementary particles.
Let us consider the case of light. Photons are $U(1)$ gauge bosons and they
cannot correspond to the breaking of any local symmetry \cite{wenlight}. Nevertheless, they can be collective modes of a different kind
of order, and this is the case of topological order. Indeed models with
topological order can feature photons, fermions and even gravitons as
emerging collective phenomena \cite{wentop,gravity}.

Light emerges from topological order as the effective low-energy theory of a
quantum spin system. The quantum spin system is built as a \emph{local
bosonic model}, namely a system in which the principle of locality is
enforced by the fact that the Hilbert space decomposes in a direct product
of local Hilbert spaces and all the observables have to commute when
far apart. Moreover, the Hamiltonian must be a sum of local observables. In
the low-energy sector, and in the continuum limit, the effective theory can
be described by the Lagrangian of electromagnetism. Therefore low-energy
excitations behave like photons. Maybe this is what photons really are,
collective excitations of a spin system on a lattice with Planck-scale
distance. But then, why do we not  see signals that are faster than light?
There could be all sorts of interactions that can propagate
as fast as permitted by the coupling constants of the underlying spin model.
A theory of light as an emergent phenomenon needs to explain why we do not see signals faster than light.

In this letter, we exploit the Lieb-Robinson bounds to show that
the maximum speed of the interactions is of the same order of magnitude than
the speed $c$ of emerging light. This answers why we can think of light as an
emergent phenomenon and still not see any faster signals in this model. In the last part
of the paper, we argue that the maximum speed increases linearly with
the dimension of the space and consider the  implications for the
horizon problem in cosmology.

\emph{Topological Order and Artificial Light.---} If we want to impose the
principle of locality in a strong sense, we must consider \emph{local
bosonic models} \cite{wenlight}. Fermionic models are not really local
because fermionic operators do not generally commute even at distance. A
local bosonic model is a theory where the total Hilbert space is the tensor
product of local Hilbert spaces, local physical operators are finite
products acting on nearby local Hilbert spaces, and the Hamiltonian is a sum
of local physical operators. Thus local physical operators must commute when
they are far apart. If we restrict ourselves to the case of a discrete
number of degrees of freedom and finite-dimensional local Hilbert spaces, we
have a \emph{quantum spin model}. A quantum spin model can be therefore
defined as follows. To every vertex $x$ in a graph $G$ we associate a finite
dimensional Hilbert space $\mathcal{H}_{x}$. The total Hilbert space of the
theory is $\mathcal{H}=\otimes _{x\in G}\mathcal{H}_{x}$. To every finite
subset of vertices $X\subset G$, we associate the local physical operators
with support in $X$ as the algebra $\mathcal{B}(\mathcal{H}_{X})$ of the
bounded linear operators over the Hilbert space $\mathcal{H}_{X}=\otimes
_{x\in X}\mathcal{H}_{x}$. The Hamiltonian will have the form $%
H_{local}=\sum_{X\subset G}\Phi _{X}$, where to every finite subset $%
X\subset G$ we associate an hermitian operator $\Phi _{X}$ with support in $X
$. An example of local bosonic model is given by a spin $1/2$ system on a
lattice. To every vertex $x$ in the lattice we associate a local Hilbert
space $\mathcal{H}_{x}\cong \mathbb{C}^{2}$. Local physical operators are
finite tensor products of the Pauli matrices at every vertex.
\begin{figure}[tbp]
\begin{equation*}
\leavevmode\hbox{\epsfxsize=6.5 cm
   \epsffile{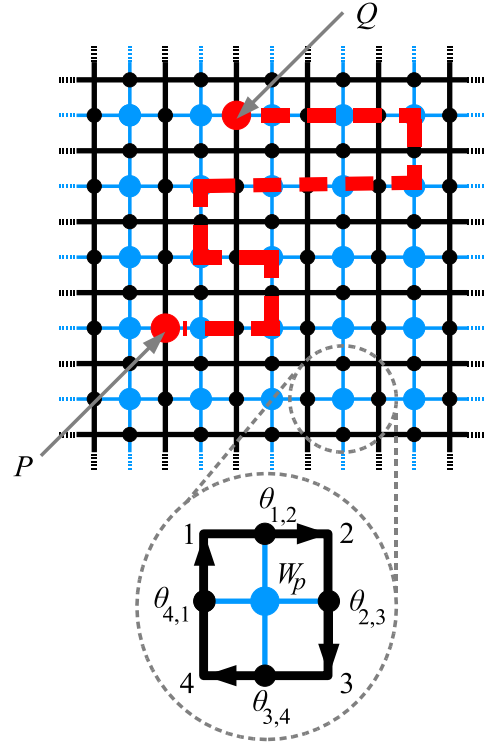}}
\end{equation*}%
\caption{(Color online) A $2D-$dimensional rotor lattice. To every plaquette $p$ is
associated a rotor operator $W_{p}$ as a function of the variables $\protect%
\theta _{ij}$. The graph $G$ is the one drawn in thin black lines. The graph $G^{\prime }$ is the graph with black and blue (lighter, bigger) dots as vertices and blue thin lines as edges. The red dashed line shows a path of length $n=22$ from the
point $P$ to the point $Q$ which are at a distance $2d(P,Q)=8$ on $G^{\prime }$ or $d(P,Q)=4$ on $G$. These paths
contain alternating link and plaquette operators.}
\label{lattice}
\end{figure}

The bosonic model we consider is a lattice of quantum rotors. Its low-energy
effective theory is a $U(1)$ lattice gauge theory whose deconfined phase
contains emergent light. Consider a square lattice whose vertices are
labeled by $\mathbf{i}$, with angular variable $\hat{\theta}_{\mathbf{ij}}$
and angular momentum $S_{\mathbf{ij}}^{z}$ on its links. The Hamiltonian for
the quantum rotor model is given by
\begin{eqnarray}
\nonumber
H_{rotor} &=&U\sum_{\mathbf{i}}\left( \sum_{\mathbf{\alpha }}S_{\mathbf{%
i,\alpha }}^{z}\right) ^{2}+J\sum_{\mathbf{i,\alpha }}(S_{\mathbf{i,\alpha }%
}^{z})^{2}  \notag \\
\nonumber
&+& \sum_{\stackrel{\mathbf{i},\{\mathbf{\alpha }_{1}, \mathbf{\alpha }_{2}\}}{ s.t. \ \mathbf{\alpha }_{1} \cdot \mathbf{\alpha }_{2} = 0}} \left( t_{<%
\mathbf{\alpha }_{1}\mathbf{\alpha }_{2}>}e^{i\left( \theta _{\mathbf{%
i+\alpha }_{1}}-\theta _{\mathbf{i+\alpha }_{2}}\right) }+h.c.\right) ,
\end{eqnarray}%
where $\mathbf{\alpha }=\pm 1/2(1,0),\pm 1/2(0,1)$ are the vectors of length $1/2$
pointing towards the lattice axes \cite{wenlight}. In the limit $%
t,J\ll U$, the first term of the Hamiltonian $H_{rotor}$ behaves like a
local constraint and makes the model a local gauge theory. Defining
$g:=2/U(t_{1 2}t_{-1 -2}+t_{2 -1}t_{-2 1})$, the effective low-energy theory becomes
\begin{equation}
H_{eff}=J\sum_{<\mathbf{ij}>}(S_{\mathbf{ij}}^{z})^{2}-g\sum_{\mathbf{p}}\frac{W_{%
\mathbf{p}}+h.c.}{2}\equiv \sum_{<\mathbf{ij}>,\mathbf{p}} (\Phi^1_{<\mathbf{ij}>} +\Phi^2_{\mathbf{p}} )\label{eff}
\end{equation}%
where $W_{\mathbf{p}}=e^{i(\theta _{12}-\theta _{23}+\theta _{34}-\theta _{41})}$
is the operator that creates a string around the plaquette $\mathbf{p}$ (see
Fig. {\ref{lattice}) and the t's are coupling constants. Although a lattice gauge theory is not a local bosonic
model, this does not violate locality because }$H_{eff}${\ is just an
effective theory. The fundamental theory is local and $H_{eff}$ is
still a sum of local terms. In the large $g/J$ limit,
the continuum theory for the Hamiltonian }$H_{eff}${\ is the Lagrangian of
electromagnetism
\begin{equation}
\nonumber
L=\int d^{2}\mathbf{x}\left( \frac{1}{4J}\mathbf{E^{2}}-\frac{g}{2}\mathbf{B}%
^{2}\right) ,
\end{equation}%
with speed of light given by $c=\sqrt{2gJ}$. 
}

\emph{Lieb-Robinson Bounds and the speed of sound in spin systems.---} Here
we review the proof of the standard Lieb-Robinson bounds \cite{lieb} in the
variant first proven in \cite{koma} and also exposed in \cite{sims}.
We consider a Hamiltonian of the form
$
H:=\sum_{X\subset G}\Phi _{X}$.
Now consider an operator $O_{Y}$ with support in a set $Y\subset G$. The
time evolution for this operator under the unitary induced by $H$ is
$O_{Y}(t)=e^{itH}O_{Y}e^{-itH}$.

The Lieb-Robinson bound is an estimate of an upper bound of the
commutator of two operators $O_{P}(t),O_{Q}(t^{\prime })$ with support in
different regions $P$ and $Q$ and at different times $t$ and $t^{\prime }$.
If the interaction map $\Phi _{X}$ couples only nearest-neighbor degrees of
freedom, the Hamiltonian can be written as $H=\sum_{<\mathbf{ij}>}h_{\mathbf{%
ij}}$ and the Lieb-Robinson bound reads
\begin{eqnarray}
\Vert \lbrack O_{P}(t),O_{Q}(0)]\Vert  &\leq &2\Vert O_{P}\Vert \Vert
O_{Q}\Vert \sum_{n=0}^{\infty }\frac{\left( 2|t|h_{max}\right) ^{n}}{n!}%
N_{PQ}(n)  \notag \\
\nonumber
&\leq &2\Vert O_{P}\Vert \Vert O_{Q}\Vert C\exp \left[ -a(d(P,Q)-vt)\right]
\end{eqnarray}%
where $h_{max}=\max_{<\mathbf{ij}>\in G}h_{\mathbf{ij}}$ and $N_{PQ}(s)$ is
the number of paths of length $s/2$ between the points $P,Q$ at distance $%
d(P,Q)$ in G \cite{hastings}. The constants $C,a,v$ have to be determined in
order to get the tightest possible bound. This bound is loose for several reasons: the crude maximization over $h_{\mathbf{ij}}$, the
overlook about the Hamiltonian's details, and the fact that all interactions are summed
in modulus instead than amplitude, so that destructive interference is not taken in account.

\emph{Lieb-Robinson Bound for the emergent $U(1)$ model.---} What do the
Lieb-Robinson bounds tell us about the model $H_{eff}$ with emergent light? Is the
maximum speed of the interactions something like the speed of the emergent
light or something completely different? As we have seen, this is of great
importance if we want to take seriously the theory of light as an emergent
phenomenon.

If we apply naively the Lieb-Robinson bounds to the Hamiltonian of the $U(1)$
lattice gauge theory, we see that the speed $v$ is
proportional to the strongest of the coupling constants, $v\propto g$. Since
light only exists in the phase $g\gg J$, we would have $v\gg\sqrt{gJ}$.  Fortunately, the bound can be made much tighter by examining the details of the Hamiltonian
and the specific way the interactions propagate. Consider the function $%
f(t):=[O_{P}(t),O_{Q}(0)]$. Then consider the set
$Z_{1}:=\{Z\subset G:[\Phi _{Z},O_{P}]\neq 0\}$, the
support of the complement of the commutant of $O_{P}$ in the set of
interactions. It turns out \cite{sims} that $f(t)$ obeys the differential equation
\begin{equation}
\nonumber
f^{\prime }(t)=-i\sum_{Z\subset Z_{1}}\left( [f(t),\Phi _{Z}(t)]+\left[
O_{P}(t),[\Phi _{Z}(t),O_{Q}(0)]\right] \right),
\end{equation}%
where $\Phi _{Z}(t)=e^{iHt}\Phi _{Z}e^{-iHt}$. From the above
equation, and using the norm-preserving property of unitary evolutions, we can establish \cite{sims} the bound
\begin{eqnarray}
\nonumber
&&\Vert \lbrack O_{P}(t),O_{Q}(0)]\Vert \leq \Vert \lbrack O_{P},O_{Q}]\Vert
\notag  \label{step1} \\
\nonumber
&&+2\Vert O_{P}\Vert \int_{0}^{|t|}\Vert \sum_{Z\subset Z_{1}}[\Phi
_{Z}(t),O_{Q}(0)]\Vert.
\end{eqnarray}%

\ignore{
Now we can apply the same method to evaluate the norm of $[\Phi
_{Z}(t),O_{Q}(0)]$ with $Z\subset Z_{1}$ and find that
\begin{eqnarray}
\nonumber
&&\Vert \lbrack \Phi _{Z\subset Z_1}(t),O_{Q}(0)]\Vert \leq \Vert
\lbrack \Phi _{Z\subset Z_1},O_{Q}]\Vert   \notag  \label{step2} \\
\nonumber
&+&2\Vert \Phi _{Z\subset Z_{1}}\Vert \int_{0}^{|t|}\Vert \sum_{Z\subset
Z_{2}}[\Phi _{Z}(t),\Phi _{Z_{1}}(t)]\Vert,
\end{eqnarray}%
where now $Z_{2}:=\{Z\subset G:[\Phi _{Z},\Phi _{Z^{\prime }\subset Z_1}]\neq 0\}$.}
 Successive iterations of the above formula yield
\begin{equation}
\Vert \lbrack O_{P}(t),O_{Q}(0)]\Vert \leq 2\Vert O_{P}\Vert \Vert
O_{Q}\Vert \sum_{n=0}^{\infty }\frac{(2|t|)^{n}}{n!}a_{n},  \label{bb}
\end{equation}%
where
\begin{equation}
a_{n}:= \sum_{Y_{i}\subset Z_{i}}\prod_{i=1}^{n}\Vert\Phi _{Y_{i}}\Vert ,
\label{an}
\end{equation}%
and we define 
$Z_{i+1}:=\{Z\subset G:[\Phi _{Z},\Phi _{Z^{\prime }\subset Z_i}]\neq 0\}$ and where $O_{1}$ and $O_{2}$ are two non-commuting local operators of the Hamiltonian.
The meaning of the above expression is the following. Every element of the
sum is a product of the type $\prod_{i}\Vert\Phi _{i}\Vert$ such that $[\Phi _{i},\Phi
_{i-1}]\neq 0$ for every $i$. If each $\Phi _{i}$ is a local bosonic
operator, every one of those products is a path on the lattice.

Let us apply these considerations to the case of the effective Hamiltonian $%
H_{eff}$. For the sake of simplicity, consider $O_{P},O_{Q}$ to be the spin
operator $S^{z}$ at the points $P,Q$. For this
Hamiltonian, the only non commuting operators are $W_{p}$ and $S_{i}^{z}$
when they have a vertex in common (see Fig.\ref{lattice}). Therefore, a path
in (\ref{an}) will consist of steps from a plaquette to any of the four links
bordering it, alternated with steps from a link to any of the two incident plaquettes.
Any such path is then a path drawn with dashed edges in Fig.\ref{lattice} on the lattice $G^{\prime }$. To every path of length $n$ on $G^{\prime }$
will then correspond an operator whose norm is $\prod_{i=1}^{n}\Vert
\Phi _{i}\Vert =(gJ)^{n/2}$. Therefore, denoting by $%
N'_{PQ}(n,d)$ the number of paths of length $n$ on $G^{\prime }$ from $P$ to a given
point $Q$ at a distance $2d$, we obtain the following bound
\begin{equation}
a_{n}\leq N'_{PQ}(n,d)(gJ)^{\frac{n}{2}} \label{cc}
\end{equation}%
A gross bound is given by $N'_{PQ}(n,d)\le 2\sqrt{8}^{n}e^{\kappa (n-2d+4)}$, for every $\kappa>0$. This is because there are 8 ways to do a succession of two steps on $G^{\prime }$: 4 choices from a blue vertex and 2 from a black one (see Fig.\ref{lattice}), and this quantity is greater than one iff there is at least one path of length $n$ between $P$ and $Q$.  Moreover, the iteration of Eq.(\ref{bb}) can be built by replacing the $2|t|$ with $\Vert \lbrack \Phi^{1},\Phi^{2}]\Vert (\Vert \Phi^{1}\Vert \Vert \Phi^{2}\Vert)^{-1} |t| = \sqrt{2} |t|$, and obtain
\begin{eqnarray}
\Vert \lbrack O_{P}(t),O_{Q}(0)]\Vert
\ignore{ & \leq & \frac{4 e^{4\kappa}}{e^{2\kappa d}}\Vert O_{P}\Vert \Vert
O_{Q}\Vert \sum_{n=0}^{\infty }\frac{(\sqrt{16 g J} e^\kappa |t|)^{n}}{n!} \nonumber \\
& = &}
\leq 4 e^{4\kappa} \Vert O_{P}\Vert \Vert
O_{Q}\Vert e^{-2\kappa(d - \frac{\sqrt{2}\sqrt{2 g J} e^\kappa }{\kappa}|t|)} . \nonumber 
\end{eqnarray}%
Optimizing for $\kappa$ we get $v_{LR}=\sqrt{2}e\sqrt{2gJ}\equiv \sqrt{2} e\times c$.
We verify numerically that our approximation for $N'(n,d)$ is good: An exact combinatorial formula is (we drop the subscript $PQ$ from now on):
\begin{equation}
\nonumber
N'(n,d)=\sum_{k=1}^{\left\lceil (n-d)/2-1\right\rceil }\binom{n-2k%
}{\frac{n-2k-d}{2}}\binom{n-2k}{\frac{n-2k}{2}}\binom{n}{2k}4^{2k}.
\end{equation}%
We numerically studied the quantity $%
\sum_{n=0}^{\infty }(\sqrt{2}|t|)^{n}a_{n}/n!$ because that is the one that enters
the bound Eq. (\ref{bb}). The factorial at the denominator makes the series
converge rapidly and we obtain, together with Eq. (\ref{bb})
\begin{equation}
\Vert \lbrack O_{P}(t),O_{Q}(0)]\Vert \leq 2\Vert O_{P}\Vert \Vert
O_{Q}\Vert Ae^{-(\frac{d-vt}{\xi })}  \label{LRS}
\end{equation}%
The speed $v$ is estimated numerically as $v \approx \sqrt{2}e\sqrt{2gJ}\equiv v_{LR} = \sqrt{2}e\times c$. Let us try to understand this result. Eq. (\ref{LRS}) establishes that all
the observables that are outside of the effective light cone centered on $P$
with speed of light $v_{LR}$ will have an exponentially small commutator
with the observables in $P$. This result sets a limit to the speed
of interactions in the spin system.  It proves that any signal outside of a light cone generated with a speed that is of the same order of magnitude (and with the same dependence on coupling constants) of light will be exponentially suppressed. We consider this result a strong indication that the maximum speed of signals is light. So the theory of emerging light explains
why its speed is also the maximum speed for any signal at low energies.
 If we were able to probe energies of order $U$ we could still find faster signals.

\emph{The cosmological horizon problem.---}
The isotropy of the cosmic microwave background presents us with the {\em horizon problem}: how is it possible that regions that
were never causally connected have the same temperature?
The horizon problem arises
from the stipulation that interactions cannot travel faster than a finite
speed, which defines a causal cone.
Inflation solves the horizon problem by introducing an exponentially fast early expansion
which allows for initial causal contact and thermalization of the observable universe  \cite{horizon}.
Alternative proposed
solutions require a mechanism for changing the speed of light as we trace
the history of the universe backwards in time \cite{maguejo} or a bimetric theory \cite{bimetric}.
Dynamically emerging light could also  resolve the horizon problem. We now wish argue that our results on the maximum speed of interactions in speed systems can be
used to justify this.

Let us first understand how the speed $v_{LR}$ depends on the dimension $%
D$ of the space. Consider a hypercubic $D-$dimensional lattice. The number
of paths of length $n$ on a hypercubic $D-$dimensional lattice will be $%
N_D(n)\sim [4 D(D-1)]^{n/2}$.
If the two dimensional case is any indication, a good enough approximation for $N_D(n,d)$, the number of paths of length n between two points $P$ and $Q$ at distance $2 d$ apart will thus be $N_D(n,d)\sim [4 D(D-1)]^{n/2}e^{\kappa(n-2d)} $, which implies a speed $v_{LR}(D)$ growing linearly with $D$
\footnote{%
Evaluating $N_D(n,d)$ exactly is a non-trivial combinatorial problem to be considered in future work.}. Now consider a model of the universe in which we start with an extremely
connected graph that evolves towards a less and less connected graph, for instance a hypercubic lattice of dimension $D(t)= D_{in}(1-\alpha t)$. This type of situation has been hypothesized in quantum spin models of the universe
like quantum graphity \cite{qgraphity}. In such a system, the
maximum speed of interactions will decrease linearly in 
with the dimension $D$ of the space, and hence in time, providing a possible explanation to the horizon problem in cosmology. Light cones then have parabolic sides and allow correlations at early times without violating causality since the distance of correlated points is of the order of $\alpha (t_i-t_f)^2$.

\emph{Conclusions.---} In this letter, we applied the technique of the
Lieb-Robinson bounds to estimate the maximum speed of interactions for a
quantum spin model with topological order and emerging $U(1)$ gauge
symmetry. The importance of this model is that the low energy excitations
are photons. Light can be regarded as an emergent phenomenon and photons can
be seen as collective modes instead of elementary particles \cite{wenlight}%
. This theory poses the problem of why we do not see other excitations that are faster
than light. The technique of the Lieb-Robinson
bounds, in the variation presented here, shows that the maximum speed of
excitations in the model has the same order of magnitude as the speed of light. Of course, it is easy to construct a different model where there is emerging light {\em and} other faster particles. One of the fundamental questions of physics is to explain why this does not seem to happen in nature. In order to address this question, the Lieb-Robinson technique could prove useful if it could be modified in order to find a tight bound for a frustrated model, where destructive interference prohibits other signals potentially faster than light.
In perspective, we think that this technique can prove useful to find exact results in $2D$ condensed matter models where there is scarcity of results that are not just numerical.

Finally, we have discussed the implications of  the finite speed of signals for cosmology and the horizon problem.

\emph{Acknowledgments.---} We acknowledge discussions with X.-G. Wen and B. Swingle and D. Abasto.
Research at Perimeter Institute for Theoretical
Physics is supported in part by the Government of Canada through NSERC and
by the Province of Ontario through MRI. This project was partially supported
by a grant from the Foundational Questions Institute (fqxi.org), a grant
from xQIT at MIT.


\begin{thebibliography}{99}


\bibitem{hastings} S.~Bravyi, M. B.~Hastings, and F.~Verstraete, \prl {\bf
97}, 050401 (2006)

\bibitem{lieb} E.~H.~Lieb, and D.~W.~Robinson, Comm. Math. Phys. \textbf{28},
251 (1972).

\bibitem{sims} B.~Nachtergaele, Y.~Ogata, and R.~Sims, J. Stat. Phys.
\textbf{124}, 1 (2006).

\bibitem{koma} M.B.~Hastings, and T.~Koma, Comm. Math. Phys. \textbf{265},
781 (2006); B.~Nachtergaele and R.~Sims, Comm. Math. Phys. \textbf{265},
119 (2006).

\bibitem{eisert} J.~Eisert, and T.J.~Osborne \prl {\bf 97}, 150404 (2006).

\bibitem{wentop} {X.-G. Wen}, \emph{Quantum Field Theory of Many-Body
Systems}, (Oxford Univ. Press, Oxford, 2004); Phys. Rev. B \textbf{40}, 7387 (1989); Int.
J. Mod. Phys. B \textbf{4}, 239 (1990); Adv. Phys. \textbf{44}, 405 (1995).


\bibitem{Wen:90} X.-G. Wen and Q. Niu, Phys. Rev. B \textbf{41}, 9377
(1990).

\bibitem{kitaev} {A.Y.~Kitaev} Annals Phys. \textbf{303}, 2 (2003).

\bibitem{wenlight} X.-G.~Wen, \prd {\bf 68}, 065003 (2003); \prb {\bf 68}, 115413 (2003).

\bibitem{gravity} Z.-C.~Gu, and X.-G.~Wen, arXiv:gr-qc/0606100v1


\bibitem{horizon} A.H.~Guth, Phys. Rev. D {\bf 23}, 347 (1981); A.~Linde, Phys. Lett. B {\bf 108}, 389 (1982).

\bibitem{maguejo} J.~Magueijo, Rep. Prog. Phys. {\bf 66}, 2025 (2003).

\bibitem{bimetric} J.~Moffat, Int. J. Mod. Phys. A20 (2005) 1155-1162.

\bibitem{qgraphity} T.~Konopka, F.~Markopoulou, S.~Severini, Phys. Rev. D
\textbf{77}, 104029 (2008).

\end{thebibliography}
\end{document}